\documentclass[sigconf, nonacm]{acmart}

\newcommand\vldbdoi{XX.XX/XXX.XX}

\newcommand\vldbavailabilityurl{URL_TO_YOUR_ARTIFACTS}

\usepackage{xcolor}
\usepackage{subcaption}
\usepackage{multirow}
\usepackage{listings}
\usepackage{algpseudocode}
\usepackage{enumitem}

\usepackage[linesnumbered,ruled,vlined]{algorithm2e} 

\begin{document}
\title{Guided Table Retrieval for Structured Data Search}

\author{Alekh Jindal,
Jyoti Pandey,
Christina Pavlopoulou,
Ronith PR,
Sharath Prakash,
Shi Qiao,
Shivani Tripathi,
Wangda Zhang}
\email{research@tursio.ai}
\affiliation{%
  \vspace{0.2cm}
  \institution{Tursio}
  \city{Bellevue}
  \country{USA}
  \vspace{0.2cm}
}

\begin{abstract}
Answering natural language questions over structured databases requires identifying the relevant tables \emph{and} determining how to join them---a task that demands both schema knowledge and semantic understanding of the user's intent.
We present \emph{guided table retrieval}, a four-phase pipeline that combines deterministic grounding via hash-based predictors, structural exploration of join-graph reachability, LLM-powered disambiguation of sources and targets, and algorithmic merging into minimal, topologically ordered join trees.
By decomposing the problem into phases with distinct responsibilities---determinism, coverage, semantic reasoning, and coherence---the pipeline avoids the brittleness of end-to-end LLM approaches while leveraging LLMs where their contextual judgment is most needed.
We evaluate on BIRD-DEV and the enterprise-scale BEAVER benchmark, achieving 94\% and 70\% precision respectively, with 92\% and 53\% F1---substantially outperforming existing baselines on precision and F1 while producing exact join trees that can be directly consumed by downstream query compilers.
\end{abstract}

\maketitle



\section{Introduction}
\label{sec:introduction}

Structured databases organize data into tables connected through foreign-key relationships, forming rich join graphs that encode the semantics of the underlying domain.                                  
Retrieving relevant information from such databases requires not only identifying the right tables but also determining how to join them---a task that demands both schema knowledge and an understanding of the query's intent. This is particularly challenging for users who are unfamiliar with the schema or the query language, as they cannot be expected to specify which tables to retrieve or how to connect them.        


\begin{figure}[t]
  \vspace{0.2cm}
  \centering
  \includegraphics[width=\columnwidth]{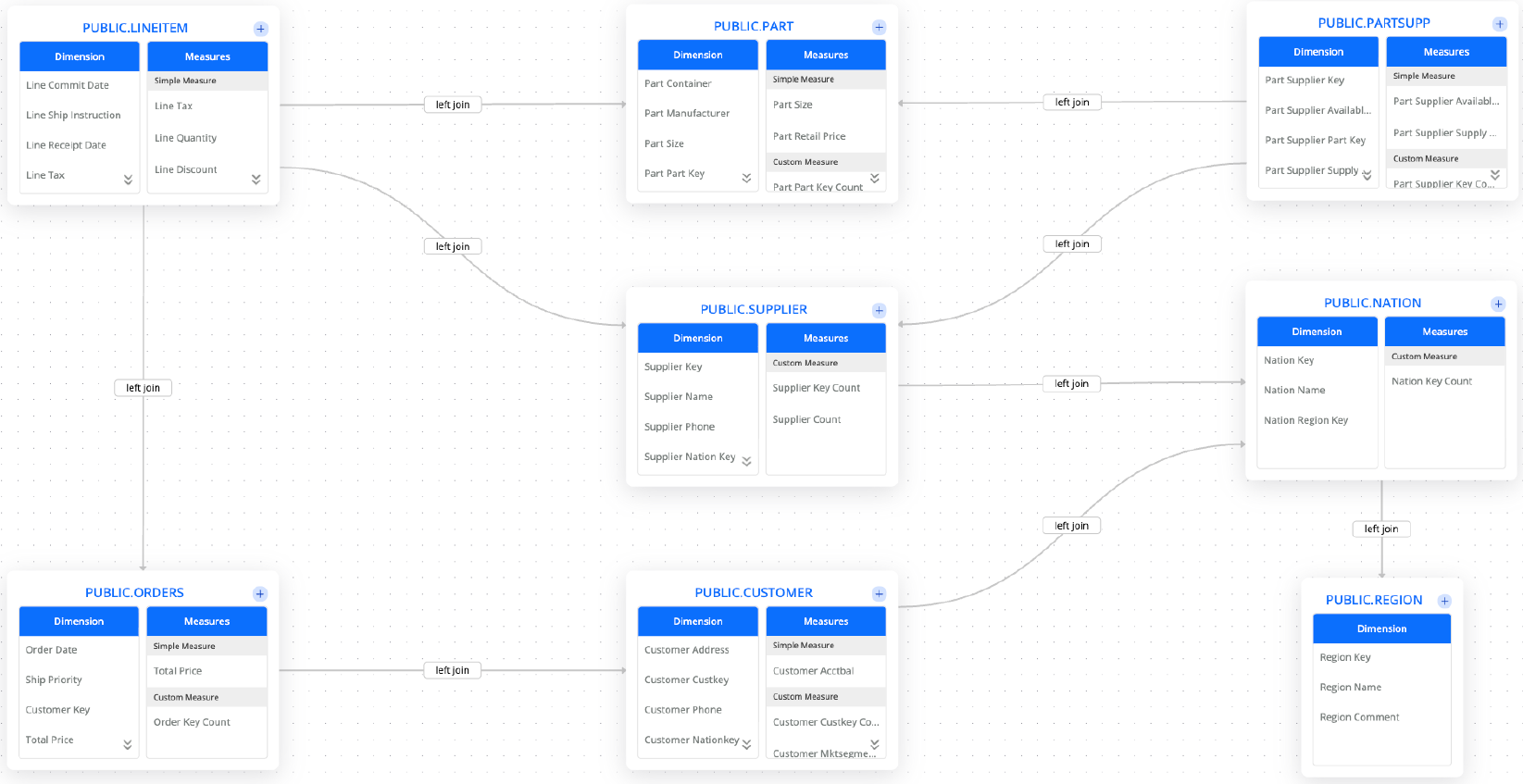}
  \caption{The TPC-H join graph. A natural-language question such as ``\emph{Which Indian suppliers are shipping the most to China?}'' requires navigating this graph to identify six tables and five joins---most of which are never mentioned explicitly in the question.}
  \vspace{-0.4cm}
  \label{fig:tpch}
\end{figure}

Consider the TPC-H benchmark~\cite{tpch_benchmark} schema shown in Figure~\ref{fig:tpch}, and the natural-language question: \emph{``Which Indian suppliers are shipping the most to China?''}
Answering this question requires the following SQL query:

\begin{lstlisting}[
  language=SQL,
  basicstyle=\ttfamily\scriptsize,
  keywordstyle=\bfseries,
  breaklines=true,
  frame=single,
  xleftmargin=1em,
  xrightmargin=1em,
  aboveskip=0.5em,
  belowskip=0.5em
]
SELECT S.S_NAME AS Supplier_Name,
       SUM(L.L_QUANTITY) AS Total_Quantity,
       COUNT(DISTINCT L.L_ORDERKEY) AS Distinct_Orders
  FROM LINEITEM L
  JOIN SUPPLIER S      ON L.L_SUPPKEY  = S.S_SUPPKEY
  JOIN NATION   SN     ON S.S_NATIONKEY = SN.N_NATIONKEY
  JOIN ORDERS   O      ON L.L_ORDERKEY  = O.O_ORDERKEY
  JOIN CUSTOMER C      ON O.O_CUSTKEY   = C.C_CUSTKEY
  JOIN NATION   CN     ON C.C_NATIONKEY = CN.N_NATIONKEY
 WHERE SN.N_NAME = 'INDIA'
   AND CN.N_NAME = 'CHINA'
 GROUP BY S.S_NAME
 ORDER BY Total_Quantity DESC;
\end{lstlisting}

\noindent Although the question mentions only suppliers and two countries, the correct query involves six tables and five joins---and reveals several non-trivial retrieval challenges:
\begin{enumerate}[leftmargin=*,nosep]
  \item The \textsc{Customer} table is needed implicitly: ``shipping to China'' refers to customer locations, not a destination column.
  \item The \textsc{Nation} table appears \emph{twice} in different roles---once for the supplier's nationality and once for the customer's---requiring contextual disambiguation.
  \item The \textsc{Lineitem} table is never mentioned by name, yet it provides the shipping metrics (quantity, order count) that are needed to answer the question.
  \item The \textsc{Orders} table serves purely as a bridge, connecting suppliers to customers through the join graph despite having no columns directly relevant to the question.
\end{enumerate}

\noindent Clearly, high-quality search over structured data must go beyond surface-level keyword matching: it needs to unpack the intent of a natural-language query into the relevant tables \emph{and} their join relationships, taking into account both semantic context and the structure of the join graph.

Existing approaches fall short on one or more of these dimensions.
Schema linking methods attempt to map question tokens to schema elements~\cite{SchemaLinking20, RESDSQL23}, but operate on individual tables without reasoning about join paths.
Retrieval-augmented generation (RAG) methods retrieve individual tables and provide them as context for LLMs to generate SQL~\cite{DinSQL23, DailSQL24, CHESS24}, but treat each table independently and ignore the join structure.
Feeding the full schema to the LLM is infeasible for large databases with hundreds of tables, as it explodes context length and cost---a challenge highlighted by enterprise benchmarks such as BEAVER~\cite{beaver_benchmark}.
Other approaches rely on sample queries to infer relevant tables and joins~\cite{VannaAI}, but are inherently brittle: they depend on the coverage and quality of the sample queries, and exhibit unpredictable behavior when the user's question falls outside the sample distribution.
Meanwhile, table discovery methods for data lakes~\cite{Josie19, Starmie23, Santos23} focus on finding unionable or joinable tables across heterogeneous sources, but do not address the problem of assembling a coherent join tree for a specific query over a known schema.

In this paper, we propose \textbf{Guided Table Retrieval}, a systematic approach that combines deterministic grounding with graph-based exploration and LLM-powered disambiguation.
Our approach proceeds in four stages:

\begin{enumerate}[leftmargin=*,nosep]
  \item \textbf{Grounding.} We use hash-based predictors to identify tokens (and their linguistic variants) in the question that match schema elements or data values. Carefully constructed, clean yet representative sample values, along with disambiguated schema elements, ensure high recall with deterministic precision.
  \item \textbf{Exploration.} Starting from the grounded tables, we enumerate reachability in the join graph to discover all source tables that can reach the grounded targets, and compute their coverage---i.e., how many target tables each source can reach.
  \item \textbf{Disambiguation.} We use LLM-based semantic inference to select the best source table and refine the set of target tables, leveraging the reachability graph and coverage information. The LLM may prune spurious targets or introduce additional ones that the grounding step missed.
  \item \textbf{Merging.} We collect the join-graph paths from the selected source to each target, compute minimal prefixes that cover all targets, and combine them into a compact, topologically ordered join tree.
\end{enumerate}

\noindent This design is deliberate in how it allocates responsibilities: grounding provides \emph{determinism}, ensuring that schema matches are precise and reproducible; exploration provides \emph{coverage}, systematically enumerating join-graph structure rather than guessing at connections; disambiguation provides \emph{semantic reasoning}, leveraging LLMs where their strength---contextual judgment---is most needed; and merging provides \emph{coherence}, synthesizing the final join tree that is both minimal and syntactically correct.
Together, these stages produce a complete join structure that can directly guide downstream SQL generation.

\section{Background}
\label{sec:background}

In this section, we describe the Tursio search platform and its foundational components, and then position table retrieval---the focus of this paper---within the broader query processing pipeline.

\subsection{The Tursio Search Platform}

Tursio~\cite{DatabasesSearchableDeepContext15} is a search platform that makes structured databases searchable in natural language.
Rather than treating natural language queries as black-box inputs to an LLM, Tursio processes them systematically using techniques inspired by traditional query planning and rewriting.
The key idea is to build a \emph{context graph}---an abstraction layer over the underlying databases---that captures table semantics, column descriptions, data profiles, join relationships, and custom measures.
Tursio then uses this context graph to contextualize user intent, compile queries, plan execution, and reason over results.
Figure~\ref{fig:architecture} illustrates the high-level architecture, which consists of four main stages: 

\begin{figure}[t]
  \centering
  \includegraphics[width=\columnwidth]{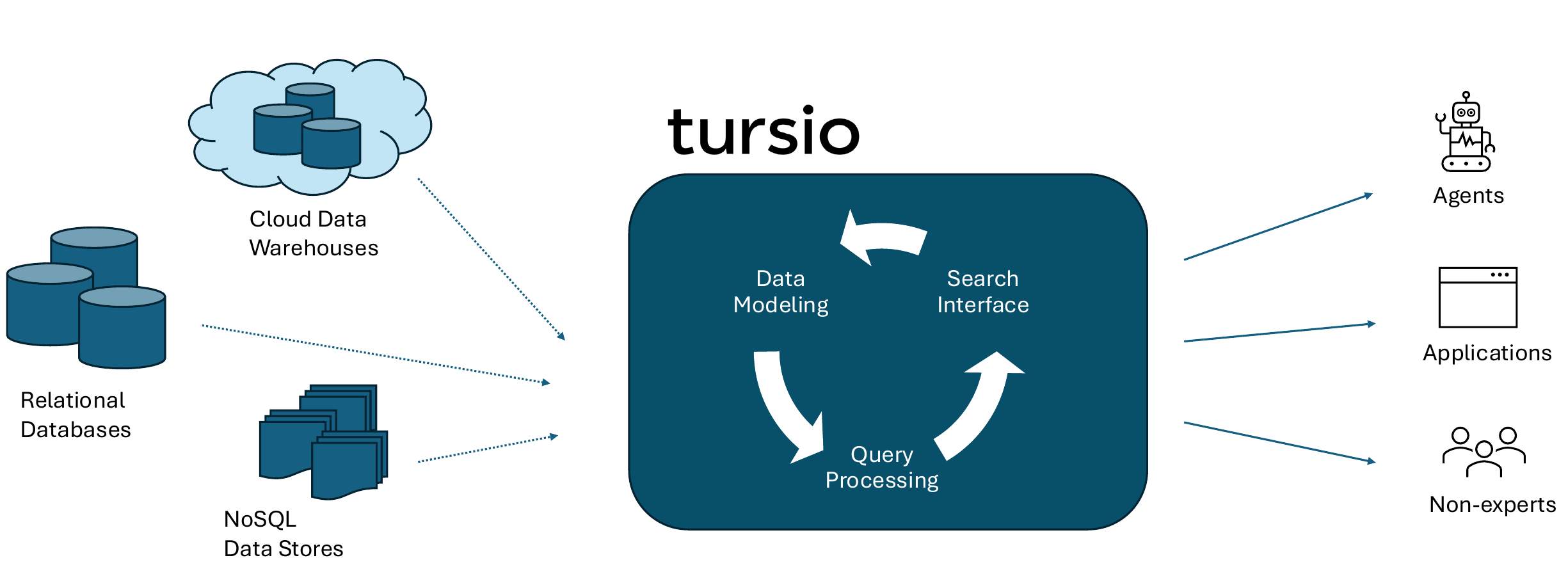}
  \caption{High-level architecture of the Tursio search platform. A context graph is inferred over the underlying databases, and natural language queries are processed through query compilation, query planning, and result reasoning---with LLMs infused into each stage.}
  \vspace{-0.2cm}
  \label{fig:architecture}
\end{figure}

\begin{enumerate}[leftmargin=*,nosep]
  \item \textbf{Data Modeling.} Tursio connects to the underlying databases and automatically infers a context graph, including table semantics, column profiles, join relationships, and custom measures.
  \item \textbf{Query Compilation.} Given a natural language question from a business user, Tursio contextualizes the user intent by identifying the relevant tables (data models) and the operations needed to answer the query.
  \item \textbf{Query Planning.} Tursio compiles the identified tables and operations into a well-formed query plan---a tree of grounded operators---and applies rewriting transformations to improve correctness and expressiveness.
  \item \textbf{Result Reasoning.} Tursio executes the query plan and presents the results with natural language summaries, visualizations, and explainability features.
\end{enumerate}

\noindent LLMs are infused into every stage of this pipeline, but each stage constrains the LLM's role to a specific, well-defined task---mirroring the philosophy from learned query optimization~\cite{jindal2021microlearner, steeringOptimizers} that decomposing into fine-grained, model-augmented steps yields better accuracy and debuggability than a single monolithic model.

\subsection{Context Graph}

The context graph is the foundation that enables Tursio to understand the database and contextualize user queries.
It is inferred automatically when a database is connected, and evolves over time as the underlying data changes.

\paragraph{Profiling.}
Tursio profiles each column by collecting fixed-size samples (100K--1M rows), applying data cleaning to remove outliers and gather representative values.
These profiles capture column types, statistics (count, distinct count, min, max), ontologies, sample values, and data types.

\paragraph{Semantics.}
Tursio uses LLMs to infer human-readable descriptions for tables and columns, resolving cryptic naming conventions common in enterprise databases (e.g., mapping \texttt{cust\_acctbal} to ``Customer Account Balance'').
These descriptions are stored as aliases in the context graph and used during query compilation to match user intent to schema elements.

\paragraph{Join Inference.}
Building an accurate context graph requires identifying how tables relate to one another.
Tripathi et al.~\cite{ScalableJoinInference26} describe a scalable approach to inferring join relationships that combines statistical pruning with LLM-based adjudication.
The approach first infers primary key candidates using lightweight statistics (count, distinct count, approximate distinct count) and heuristic scoring, then detects inclusion dependencies across table pairs using sample-based overlap analysis.
LLMs are used to adjudicate ambiguous candidates---leveraging semantic understanding of table and column names to distinguish meaningful foreign-key relationships from coincidental value overlaps.
This statistics-LLM combination scales to large schemas (hundreds of tables with hundreds of thousands of candidate pairs) while maintaining high precision.
The resulting join relationships form the edges of the context graph, connecting tables into a coherent join graph that is central to the table retrieval problem addressed in this paper.

\paragraph{Join Trees.}
The inferred join relationships are organized into \emph{left-join trees}---rooted trees where the root (fact table) has the most foreign-key references, and dimension tables are attached along their reachable join paths.
These join trees encode valid ways to combine tables without introducing data duplication, and serve as the structural backbone for query compilation.

\subsection{Search Quality Evaluation}

In our earlier work, we presented an evaluation of the Tursio search platform, comparing it against ChatGPT and Perplexity on a realistic banking schema~\cite{TursioSearchEval26}.
Their results show that Tursio achieves answer relevancy statistically comparable to both baselines (97.8\% vs.\ 98.1\% on simple, 90.0\% vs.\ 100.0\% on medium, and 89.5\% vs.\ 100.0\% on hard questions), despite Tursio answering from a structured database while the baselines generate responses from the open web.
Crucially, their analysis identifies \emph{database completeness}---not model comprehension---as the primary bottleneck.
They also highlight that business users ask high-level, abstract questions that omit explicit references to tables, join paths, or metric definitions: the ratio of SQL token-length to question token-length is significantly higher in enterprise workloads than in academic benchmarks like BIRD~\cite{bird_bench}.

This finding directly motivates the present work.
When business users ask questions like ``\emph{Which Indian suppliers are shipping the most to China?}'', the system must bridge the gap between the high-level intent and the underlying schema---identifying not just which tables are relevant, but how they connect through the join graph.
The table retrieval problem, which we formalize and solve in this paper, is the critical first step in this bridge: it takes a natural language question and produces a complete join tree that downstream query compilation can use to generate correct SQL.

\subsection{Problem Statement}

Given a natural language question $q$ and a context graph $\mathcal{G} = (\mathcal{T}, \mathcal{E})$, where $\mathcal{T}$ is the set of tables and $\mathcal{E}$ is the set of join relationships (edges), the \emph{table retrieval problem} is to identify:
\begin{enumerate}[leftmargin=*,nosep]
  \item A set of \emph{target tables} $T \subseteq \mathcal{T}$ that are relevant to answering $q$,
  \item A \emph{source table} $s \in \mathcal{T}$ from which all targets are reachable in $\mathcal{G}$, and
  \item A \emph{join tree} $J$ rooted at $s$ that connects $s$ to all tables in $T$ via edges in $\mathcal{E}$, such that $J$ is minimal (no unnecessary tables) and topologically ordered (respecting join directions).
\end{enumerate}

\begin{figure*}[t!]
  \centering
  \includegraphics[width=0.9\textwidth]{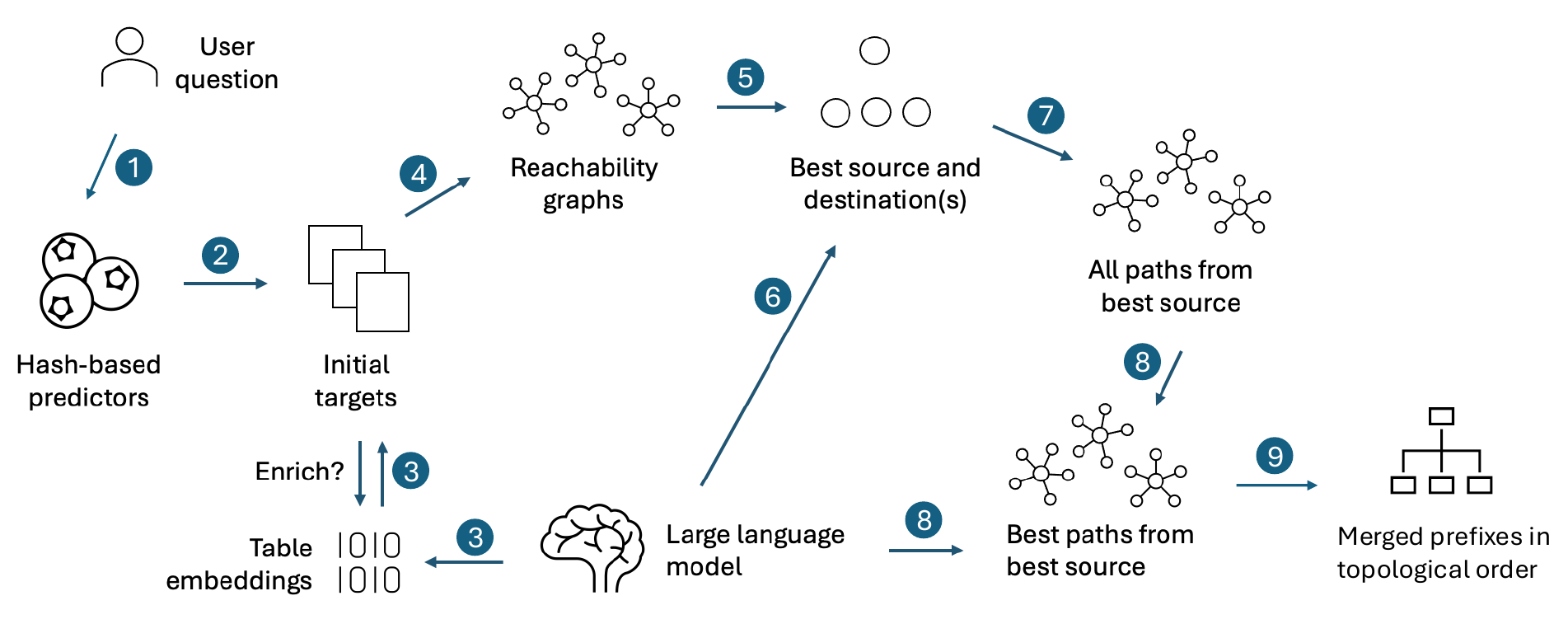}
  \vspace{-0.4cm}
  \caption{The guided table retrieval pipeline. A user question is first grounded to initial target tables via hash-based predictors (steps~1--2), optionally enriched with table embeddings (step~3). Reachability graphs are then explored to identify candidate source tables and their coverage (step~4). An LLM disambiguates the best source and destination tables (steps~5--6), and all paths from the selected source are enumerated (step~7). The LLM selects the best paths (step~8), and their prefixes are merged into a compact join tree in topological order (step~9).}
  \label{fig:table-retrieval}
  \vspace{-0.2cm}
\end{figure*}

\noindent The output join tree $J$ is then passed to the query compilation stage, which grounds the parsed operators onto the tables in $J$ and generates the final SQL query.
This decomposition---retrieving tables first, then compiling the query---allows each stage to focus on its strength: table retrieval reasons about schema structure and join-graph topology, while query compilation reasons about operators, filters, and aggregations.

\section{Guided Table Retrieval}
\label{sec:guided-table-retrieval}

Figure~\ref{fig:table-retrieval} illustrates the end-to-end guided table retrieval pipeline.
Given a natural language question and a context graph, the pipeline proceeds through four phases---\emph{grounding}, \emph{exploration}, \emph{disambiguation}, and \emph{merging}---corresponding to the nine numbered steps in the diagram.
Each phase is deliberately scoped: grounding is deterministic and fast, exploration is structural and exhaustive, disambiguation leverages LLM reasoning where it is most needed, and merging is algorithmic and guaranteed to produce a valid join tree.
We describe each phase in detail below, using the TPC-H example from Section~\ref{sec:introduction} as a running illustration.

\subsection{Grounding}
\label{sec:grounding}

The first phase identifies an initial set of \emph{target tables}---tables that are lexically or semantically connected to the user's question (steps~1--3 in Figure~\ref{fig:table-retrieval}).

\paragraph{Hash-based prediction (steps 1--2).}
We maintain an inverted index that maps tokens to schema elements and sampled data values.
Given a question $q$, we tokenize it and generate linguistic variants for each token---including lemmas, synonyms, and common abbreviations.
Each variant is hashed and looked up in the index, which stores mappings of the form:
\begin{equation*}
  \texttt{hash}(token) \;\longrightarrow\; \{(table, column, \textit{match\_type})\}
\end{equation*}
where \textit{match\_type} distinguishes between schema matches (the token matches a table name, column name, or alias) and value matches (the token matches a sampled data value in a column).
All tables that receive at least one match are collected into the initial target set of tables $T_0$.

\smallskip
\noindent\textbf{Example.} For the question ``\emph{Which Indian suppliers are shipping the most to China?}'', the hash predictor matches:
\begin{itemize}[leftmargin=*,nosep]
  \item ``Indian'' and ``India'' $\rightarrow$ value match in \textsc{Nation.N\_Name}
  \item ``suppliers'' $\rightarrow$ schema match on \textsc{Supplier} (table name)
  \item ``shipping'' $\rightarrow$ schema match on \textsc{Lineitem.L\_ShipDate} (column alias)
  \item ``China'' $\rightarrow$ value match in \textsc{Nation.N\_Name}
\end{itemize}
This yields an initial target set $T_0 = \{\textsc{Nation}, \textsc{Supplier}, \textsc{Lineitem}\}$.

\paragraph{Semantic enrichment (step 3).}
Hash-based prediction offers high precision but may miss tables that are relevant yet not lexically present in the question.
To improve recall, we apply a semantic fallback when the hash predictor returns fewer than $k$ targets (we use $k{=}3$ in practice).
Following the deep context approach of Tursio, described in our earlier work~\cite{DatabasesSearchableDeepContext15}, each table in the context graph has a precomputed embedding derived from its synthetically generated query context---a set of representative natural-language questions that the table could help answer.
We embed the user's question using the same model and retrieve the top-$k$ most similar tables by cosine similarity, adding any new tables to $T_0$.

In our running example, the hash predictor already returns three targets, so semantic enrichment is not triggered.
However, for a vaguer question like ``\emph{How are our sales trending?}'', the hash predictor might only match \textsc{Orders} (via ``sales''), and semantic enrichment would add \textsc{Lineitem} and \textsc{Customer} based on their query-context similarity.

\subsection{Exploration}
\label{sec:exploration}

The second phase discovers which tables in the context graph can serve as the \emph{source} (root) of a join tree that reaches the initial targets (step~4 in Figure~\ref{fig:table-retrieval}).

\paragraph{Reachability computation.}
For each table $t \in \mathcal{T}$, we compute its \emph{reachability set} $R(t) \subseteq \mathcal{T}$: the set of all tables reachable from $t$ by following directed join edges in the context graph.
Since join trees in Tursio are rooted left-join trees, reachability follows the direction of foreign-key relationships---from fact tables (parents) toward dimension tables (children).
The reachability sets are precomputed and cached, so this step incurs no per-query cost.

\paragraph{Coverage scoring.}
Given the initial target set $T_0$, we compute the \emph{coverage} of each table as:
\begin{equation*}
  \textit{coverage}(t) = |R(t) \cap T_0|
\end{equation*}
Tables with higher coverage can reach more of the initial targets through the join graph, making them stronger candidates for the source.
We retain all tables with $\textit{coverage}(t) \geq 1$ as candidate sources, ranked by coverage.
Note that a candidate need not cover \emph{all} initial targets at this stage: the disambiguation phase (Section~\ref{sec:disambiguation}) jointly selects the source and refines the target set, ensuring that the final source $s$ table can reach all final target tables $T$ in the context graph.

\smallskip
\noindent\textbf{Example.} In the TPC-H join graph (Figure~\ref{fig:tpch}), the reachability sets of selected tables are:
\begin{itemize}[leftmargin=*,nosep]
  \item $R(\textsc{Lineitem}) = \{\textsc{Part}, \textsc{PartSupp}, \textsc{Supplier}, \textsc{Nation}, \textsc{Region}, \\\textsc{Orders}, \textsc{Customer}\}$
  \item $R(\textsc{Supplier}) = \{\textsc{Nation}, \textsc{Region}\}$
  \item $R(\textsc{Orders}) = \{\textsc{Customer}, \textsc{Nation}, \textsc{Region}\}$
\end{itemize}
Against $T_0 = \{\textsc{Nation}, \textsc{Supplier}, \textsc{Lineitem}\}$, the coverage scores are:
\textsc{Lineitem}~$= 2$ (reaches \textsc{Supplier} and \textsc{Nation}),
\textsc{Supplier}~$= 1$ (reaches \textsc{Nation}),
\textsc{Orders}~$= 1$ (reaches \textsc{Nation}).
\textsc{Lineitem} emerges as the top candidate source, since it has the highest coverage and is itself a target.

\subsection{Disambiguation}
\label{sec:disambiguation}

The third phase uses an LLM to make two semantic decisions that require understanding the user's intent: selecting the best source table and refining the set of target tables (steps~5--8 in Figure~\ref{fig:table-retrieval}).
This is where the pipeline transitions from deterministic, structure-based reasoning to LLM-powered contextual judgment.

\paragraph{Source and target selection (steps 5--6).}
We present the LLM with:
\begin{enumerate}[leftmargin=*,nosep]
  \item The user's question $q$.
  \item The candidate source tables, ranked by coverage, along with their reachability sets.
  \item The initial target set $T_0$ from the grounding phase.
\end{enumerate}
The LLM is asked to (a)~select the single best source table $s$, and (b)~refine the initial targets into the final target set $T$ (corresponding to the target tables in the problem statement of Section~\ref{sec:background}).
The refinement may \emph{prune} spurious targets---tables that matched lexically but are not semantically relevant---or \emph{extend} the set with tables that the grounding phase missed but that the LLM recognizes as necessary from context.

\smallskip
\noindent\textbf{Example.} Given the candidate sources and initial targets from the previous phase, the LLM selects \textsc{Lineitem} as the source $s$ (highest coverage, central to ``shipping'' semantics) and refines the target set to $T = \{\textsc{Supplier}, \textsc{Nation}, \textsc{Orders}, \textsc{Customer}\}$.
Notably, the LLM \emph{adds} \textsc{Orders} and \textsc{Customer}---recognizing that ``shipping to China'' implies customer locations, which requires traversing through \textsc{Orders} to reach \textsc{Customer} and then \textsc{Nation}.

\paragraph{Path enumeration (step 7).}
Once the source $s$ is selected, we enumerate all join paths from $s$ to each target $d \in T$ in the context graph.
A path $P(s, d) = [s, t_1, t_2, \ldots, d]$ is a sequence of tables connected by join edges.
When multiple paths exist between $s$ and a target (e.g., through different intermediate tables), all are enumerated for the LLM to choose from.

\paragraph{Path selection (step 8).}
The LLM is presented with all enumerated paths and asked to select the best path for each target, based on semantic relevance to the question.
This step is critical when the join graph contains multiple routes between two tables---the LLM must determine which route carries the correct semantics.

\smallskip
\noindent\textbf{Example.} From \textsc{Lineitem} to \textsc{Nation}, there are two paths in the TPC-H graph:
\begin{enumerate}[leftmargin=*,nosep]
  \item \textsc{Lineitem} $\rightarrow$ \textsc{Supplier} $\rightarrow$ \textsc{Nation} (supplier's country)
  \item \textsc{Lineitem} $\rightarrow$ \textsc{Orders} $\rightarrow$ \textsc{Customer} $\rightarrow$ \textsc{Nation} (customer's country)
\end{enumerate}
The question requires \emph{both} paths: the first to filter Indian suppliers, the second to filter Chinese customers.
The LLM recognizes this and selects both, assigning each the appropriate semantic role.
This is precisely the kind of disambiguation that purely structural methods cannot perform---it requires understanding that ``India'' and ``China'' refer to different entities in the question.

\subsection{Merging}
\label{sec:merging}

The final phase combines the selected paths into a single, compact join tree $J$ that can be passed to the query compiler (step~9 in Figure~\ref{fig:table-retrieval}).

\paragraph{Prefix extraction.}
Each selected path $P(s, d_i)$ from the source to a target corresponds to a \emph{prefix} of the source table's full join tree in the context graph.
A prefix is the minimal subtree rooted at $s$ that includes all tables on the path.
Formally, for a path $P = [s, t_1, \ldots, d_i]$, its prefix is $\textit{prefix}(P) = \{s, t_1, \ldots, d_i\}$ along with the edges connecting consecutive tables.

\paragraph{Prefix merging.}
When multiple paths share intermediate tables, their prefixes overlap.
We merge all prefixes by taking their union, deduplicating shared nodes and edges.
The key insight is that we only need to include tables up to the \emph{last target} on each path: if a path passes through tables beyond the final target that the LLM selected, those trailing tables are trimmed.

\smallskip
\noindent\textbf{Example.} Suppose the LLM selected source \textsc{A} and two paths in a hypothetical schema:
\begin{align*}
  P_1 &= [\textsc{A}, \textsc{B}, \textsc{C}, \textsc{D}] \\
  P_2 &= [\textsc{A}, \textsc{C}, \textsc{E}, \textsc{F}]
\end{align*}
If the final targets are $T = \{\textsc{D}, \textsc{F}\}$, both paths are needed in full, yielding the merged set $\{\textsc{A}, \textsc{B}, \textsc{C}, \textsc{D}, \textsc{E}, \textsc{F}\}$.
However, if the targets are $T = \{\textsc{B}, \textsc{F}\}$, path $P_1$ can be trimmed to $[\textsc{A}, \textsc{B}]$ (since \textsc{C} and \textsc{D} are beyond the target), yielding the smaller merged set $\{\textsc{A}, \textsc{B}, \textsc{C}, \textsc{E}, \textsc{F}\}$.

\paragraph{Topological ordering.}
The merged set of tables and edges forms a DAG rooted at the source $s$.
We compute a topological ordering of this DAG, which determines the join sequence: the source table is listed first, followed by each subsequent table in an order that respects the join dependencies (a table appears only after all tables it depends on).
The result is the join tree $J$---rooted at $s$, connecting all tables in $T$, minimal, and topologically ordered---satisfying all three requirements of the problem statement in Section~\ref{sec:background}.

\smallskip
\noindent\textbf{Example.} Returning to the running TPC-H example, the merged join tree is:
\begin{center}
\textsc{Lineitem} $\rightarrow$ \textsc{Supplier} $\rightarrow$ \textsc{Nation$_{\text{S}}$} $\rightarrow$ \textsc{Orders} $\rightarrow$ \textsc{Customer} $\rightarrow$ \textsc{Nation$_{\text{C}}$}
\end{center}
where \textsc{Nation} appears twice with different roles (supplier nation and customer nation), and the topological order ensures that each join can reference the tables it depends on.
This join tree is passed directly to the query compiler, which grounds the filter predicates (``India'' on \textsc{Nation$_{\text{S}}$}, ``China'' on \textsc{Nation$_{\text{C}}$}), aggregation (sum of quantity from \textsc{Lineitem}), and grouping (by \textsc{Supplier.S\_Name}) to produce the final SQL query.

\section{Evaluation}
\label{sec:evaluation}

We evaluate the guided table retrieval pipeline described in Section~\ref{sec:guided-table-retrieval} on two established benchmarks: BIRD-DEV~\cite{bird_bench} and BEAVER~\cite{beaver_benchmark}.
Our evaluation builds on the experimental setup from our earlier work on making databases searchable~\cite{DatabasesSearchableDeepContext15}, which evaluated the full Tursio search platform end-to-end; here we focus specifically on the table retrieval component and analyze its behavior in detail.

\subsection{Metrics}

Following the evaluation methodology of BEAVER benchmark~\cite{beaver_benchmark}, we measure table retrieval quality using four metrics:
\begin{itemize}[leftmargin=*,nosep]
  \item \textbf{Precision} --- fraction of retrieved tables that are in the gold set.
  \item \textbf{Recall} --- fraction of gold tables that are successfully retrieved.
  \item \textbf{F1} --- the harmonic mean of precision and recall.
  \item \textbf{Perfect Recall (PR)} --- fraction of queries for which \emph{all} gold tables are retrieved.
\end{itemize}
Perfect Recall is particularly important in production: missing even a single table typically produces syntactically valid but semantically incorrect SQL, since the query compiler cannot compensate for a missing join path.
We also report \textbf{Accuracy}---the fraction of queries where the retrieved set exactly matches the gold set---which jointly penalizes both missing and extraneous tables.

\subsection{Benchmarks}

\paragraph{BIRD-DEV}
The BIRD benchmark~\cite{bird_bench} contains 1,534 natural-language questions over 11 databases spanning diverse domains (financial, educational, sports, community, etc.).
Each database has a moderate schema (5--15 tables) with well-defined foreign keys, making it a good testbed for evaluating grounding and join-graph exploration.

\paragraph{BEAVER}
The BEAVER benchmark~\cite{beaver_benchmark} targets enterprise-scale table retrieval, with 6 datasets drawn from real-world data warehouses (DW, Keystone, CSAIL\_Stata variants).
These schemas are significantly larger (up to hundreds of tables) and feature noisier metadata, ambiguous naming conventions, and complex join graphs---precisely the conditions where guided table retrieval is most needed.
BEAVER reports baseline performance using the best results across multiple retrieval methods (embedding-based top-5 and top-10), providing a strong comparison point.

\subsection{Results on BIRD-DEV}

\begin{figure}[t]
  \centering
  \includegraphics[width=\columnwidth]{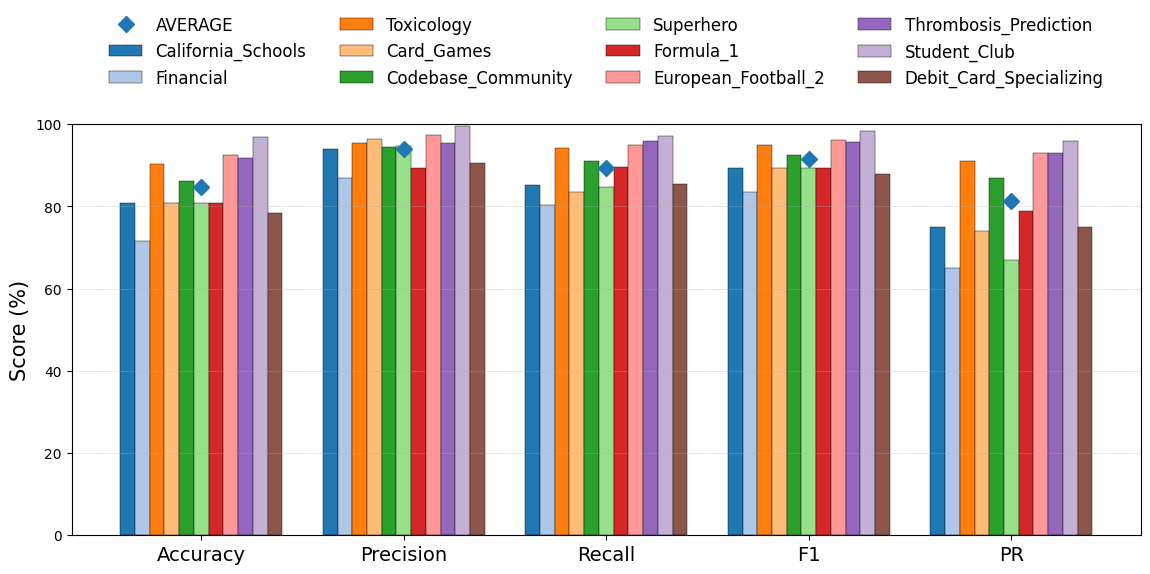}
  \caption{Table retrieval results on BIRD-DEV, broken down by database. Bars show per-database scores; the leftmost group (AVERAGE) shows the overall mean. Tursio achieves 85\% accuracy, 94\% precision, 89\% recall, 92\% F1, and 81\% Perfect Recall across all 11 databases.}
  \label{fig:table_retrieval_bird}
  \vspace{-0.4cm}
\end{figure}

Figure~\ref{fig:table_retrieval_bird} shows the table retrieval results on BIRD-DEV, broken down by database.
Tursio achieves strong overall performance: 85\% accuracy, 94\% precision, 89\% recall, 92\% F1, and 81\% Perfect Recall.
Several observations stand out.
First, \emph{precision is consistently high across all databases} (90--100\%), confirming that the grounding phase (Section~\ref{sec:grounding}) rarely introduces spurious tables---the hash-based predictors and semantic enrichment are effective at identifying genuinely relevant tables.
Second, \emph{recall varies more across databases}, reflecting differences in schema complexity and the degree to which questions implicitly reference tables.
Databases like \texttt{California\_Schools} and \texttt{Financial} achieve near-perfect recall, while more complex schemas like \texttt{European\_Football\_2} show lower recall due to indirect table references that require deeper join-graph traversal.
Third, \emph{Perfect Recall at 81\%} means that for four out of five queries, the pipeline retrieves the complete set of gold tables---a strong result given that the pipeline must navigate join graphs and resolve ambiguous references without any schema-specific tuning.

\subsection{Results on BEAVER}

\begin{figure}[t]
  \centering
  \includegraphics[width=\columnwidth]{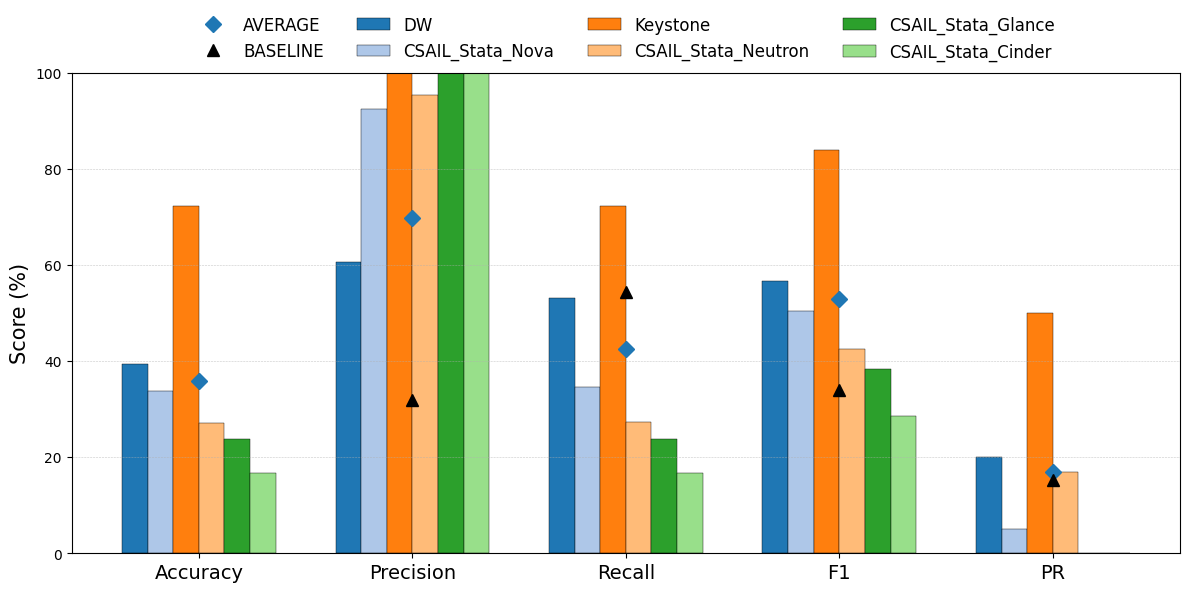}
  \caption{Table retrieval results on BEAVER, broken down by dataset. Diamond markers show the Tursio average; triangle markers show the best BEAVER baseline (best of top-5 and top-10 embedding-based retrieval across all models evaluated in~\cite{beaver_benchmark}). Tursio outperforms the baseline on precision and F1.}
  \label{fig:table_retrieval_beaver}
  \vspace{-0.4cm}
\end{figure}

Figure~\ref{fig:table_retrieval_beaver} shows the results on the BEAVER benchmark, comparing Tursio (bars) against the best-performing BEAVER baseline (triangle markers).
The baseline represents the best results across all models evaluated in the BEAVER paper~\cite{beaver_benchmark}, including both top-5 and top-10 embedding-based retrieval methods.

Tursio achieves substantially higher precision (70\% vs.\ 32\% baseline average) and F1 (53\% vs.\ 34\%), while showing comparable or slightly lower recall (43\% vs.\ 54\%).
This precision--recall trade-off is by design: unlike baseline methods that retrieve a fixed-size superset of candidate tables (top-$k$), guided table retrieval produces the \emph{exact} set of tables predicted to be necessary for the query.
This yields fewer false positives at the cost of occasionally missing tables that require deeper semantic reasoning.

The results vary across datasets, reflecting the diversity of schema characteristics in BEAVER:
\begin{itemize}[leftmargin=*,nosep]
  \item \textbf{Keystone} achieves the highest scores across all metrics (73\% accuracy, 100\% precision, 73\% recall, 84\% F1), suggesting a well-structured schema where Tursio provides strong guidance.
  \item \textbf{DW} shows high precision (93\%) but moderate recall (53\%), indicating that the grounding phase successfully identifies relevant tables but the exploration and disambiguation phases miss some implicit references in this large schema.
  \item \textbf{CSAIL\_Stata variants} show lower overall performance, reflecting schemas with particularly ambiguous metadata and complex join topologies that challenge all methods (including baseline).
\end{itemize}

\subsection{Discussion}

The evaluation reveals three key insights about Tursio's guided table retrieval.

\paragraph{Precision is the pipeline's strength.}
Across both benchmarks, Tursio consistently achieves high precision---94\% on BIRD-DEV and 70\% on BEAVER (vs.\ 32\% baseline).
This is a direct consequence of the four-phase design: grounding provides deterministic, high-confidence matches; exploration constrains the search to structurally valid join paths; disambiguation uses LLM reasoning to prune spurious candidates; and merging produces a minimal join tree with no extraneous tables.

\paragraph{Recall depends on schema complexity.}
On well-structured schemas (BIRD-DEV, Keystone), recall is high (89\% and 73\% respectively).
On noisier, larger schemas (CSAIL\_Stata variants), recall drops---primarily because implicit table references require semantic reasoning that goes beyond what the current grounding and exploration phases can capture.
Improving recall on enterprise-scale schemas is an active area of work, potentially through richer profiling, domain-specific embeddings, or multi-hop exploration strategies.

\paragraph{Exact retrieval vs.\ top-$k$ retrieval.}
A fundamental distinction between guided table retrieval and baseline methods is the retrieval granularity.
Baseline methods retrieve a ranked list of $k$ candidate tables and rely on downstream components to filter; guided table retrieval produces the exact set of tables needed, assembled into a join tree.
This makes the pipeline's output directly usable by the query compiler---no further filtering or join inference is needed---but it also means that every missed table directly impacts recall, whereas top-$k$ methods absorb misses in their larger candidate set.

\section{Related Work}
\label{sec:related-work}

We discussed table discovery in data lakes and RAG-based text-to-SQL methods in Section~\ref{sec:introduction}. Here we cover other related work.

\paragraph{Schema encoding for text-to-SQL}
Several text-to-SQL parsers encode the database schema as a graph and use neural architectures to select relevant tables and columns.
RAT-SQL~\cite{RATSQL20} uses relation-aware transformers to jointly reason over question tokens and schema elements via a schema graph.
BRIDGE~\cite{BRIDGE20} anchors database cell values in the question to improve cross-domain schema grounding.
These methods learn table selection implicitly as part of end-to-end SQL generation, whereas our approach performs table retrieval as an explicit, interpretable pipeline stage---producing a join tree that is independent of the downstream query compiler.

\paragraph{Natural language interfaces and intent disambiguation.}
NaLIR~\cite{NaLIR14} pioneered interactive natural language interfaces that construct parse trees and engage users in disambiguation dialogues to resolve ambiguous database queries.
Wolfson et al.~\cite{QDMR20} decompose complex questions into structured sub-steps (QDMR), providing an intermediate representation that can guide multi-table query formulation.
Our disambiguation phase serves a similar purpose---resolving which tables a question refers to---but does so non-interactively via a single LLM call over the reachability graph, avoiding the need for user interaction.

\paragraph{Knowledge graphs and ontology-based data access.}
Aurum~\cite{Aurum18} builds enterprise knowledge graphs over schema metadata (join paths, containment, semantic similarity) to enable graph-based data discovery.
Ontop~\cite{Ontop17} maps relational schemas to ontologies, allowing users to query through a knowledge graph abstraction that hides join complexity.
Our context graph shares the motivation of making schema structure navigable, but differs in that it is automatically inferred (via statistical profiling and LLM adjudication~\cite{ScalableJoinInference26}) and used as a runtime structure for per-query table retrieval rather than a static query interface.

\section{Conclusion}
\label{sec:conclusion}

We presented guided table retrieval, a systematic approach to identifying the tables and join paths needed to answer natural language questions over structured databases.
The pipeline decomposes the problem into four phases---grounding, exploration, disambiguation, and merging---each with a well-defined role: hash-based predictors provide deterministic anchoring, join-graph traversal ensures structural coverage, LLM reasoning resolves semantic ambiguities, and prefix merging produces minimal join trees.

Our evaluation on BIRD-DEV and BEAVER demonstrates that this decomposition achieves high precision (94\% and 70\%) and strong F1 (92\% and 53\%), substantially outperforming existing baselines.
Unlike top-$k$ retrieval methods, guided table retrieval produces exact join trees that are correct and complete, and are directly consumable by query compilers, eliminating the need for downstream table filtering or join inference.

Several directions remain open.
There is room to improve recall on enterprise-scale schemas with noisy metadata through richer profiling, domain-specific embeddings, and multi-hop exploration.
The disambiguation phase currently relies on a single LLM call; iterative refinement with feedback from the query compiler could further improve accuracy.
Finally, extending the pipeline to handle multi-turn conversations---where context accumulates across questions---is a natural next step toward making databases truly searchable for business users.
\vspace{-0.1cm}


\balance
\bibliographystyle{ACM-Reference-Format}
\bibliography{references}

\end{document}